\documentclass[conference]{IEEEtran}
\IEEEoverridecommandlockouts
\usepackage{cite}
\usepackage{amsmath,amssymb,amsfonts}
\usepackage{algorithmic}
\usepackage{graphicx}
\usepackage{textcomp}
\usepackage{xcolor}
\usepackage{pgfplots}
\pgfplotsset{compat=newest}
\usepackage{fancyhdr}
\usepackage{multirow}
\usepackage{comment}
\usepackage{textcomp}
\usepackage{graphicx}
\usepackage{multirow}
\usepackage{caption,tabularx,booktabs}

\def\BibTeX{{\rm B\kern-.05em{\sc i\kern-.025em b}\kern-.08em
    T\kern-.1667em\lower.7ex\hbox{E}\kern-.125emX}}
\begin{document}

\title{SGX-MR: Regulating Dataflows for Protecting Access Patterns of Data-Intensive SGX Applications}

\author{\IEEEauthorblockN{A K M Mubashwir Alam}
\IEEEauthorblockA{
\textit{Marquette University}\\
mubashwir.alam@marquette.edu}
\and
\IEEEauthorblockN{Sagar Sharma}
\IEEEauthorblockA{
\textit{HP Inc.}\\
sagar.shamra@hp.com}
\and
\IEEEauthorblockN{Keke Chen}
\IEEEauthorblockA{
\textit{Marquette University}\\
 keke.chen@marquette.edu}

}

\maketitle
\let\thefootnote\relax\footnotetext{*To appear in 21st Privacy Enhancing Technologies Symposium (PETS'2021)}

	\begin{abstract}{
		
		Intel SGX has been a popular trusted execution
		environment (TEE) for protecting the integrity
		and confidentiality of applications running on untrusted
		platforms such as cloud. However, the access patterns
		of SGX-based programs can still be observed by
		adversaries, which may leak important information for
		successful attacks. Researchers have been experimenting
		with Oblivious RAM (ORAM) to address the privacy
		of access patterns. ORAM is a powerful low-level
		primitive that provides application-agnostic protection
		for any I/O operations, however, at a high cost. We find
		that some application-specific access patterns, such as
		sequential block I/O, do not provide additional information
		to adversaries. Others, such as sorting, can be
		replaced with specific oblivious algorithms that are more
		efficient than ORAM. The challenge is that developers
		may need to look into all the details of application-specific
		access patterns to design suitable solutions,
		which is time-consuming and error-prone. In this paper,
		we present the lightweight SGX based MapReduce (SGX-MR) approach that regulates the dataflow of data-intensive SGX applications for easier application-level access-pattern analysis and
		protection. It uses the MapReduce framework to cover a
		large class of data-intensive applications, and the entire framework can be implemented with a small memory
		footprint. With this framework, we have examined the
		stages of data processing, identified the access patterns
		that need protection, and designed corresponding efficient
		protection methods. Our experiments show that SGX-MR
		based applications are much more efficient than the
		ORAM-based implementations.
	}
\end{abstract}

\begin{IEEEkeywords}
SGX-based data analytics, MapReduce, Data flow regularization, Access Patterns, ORAM
\end{IEEEkeywords}

	\section{Introduction}
With the development of resource-starving applications in big data, artificial intelligence, and the Internet of Things (IoT), cloud computing has become popular for its storage, computation scalability, and accessibility. However, when uploading data and conducting computations in the cloud, data owners lose full control of their data and must trust that service providers can take care of security well.  In practice, due to the wide attack surface of the cloud software stack and the unique co-tenancy feature \cite{ristenpart09}, we have witnessed frequent cloud security incidents. Practitioners using public clouds have experienced at least one cloud-related security incident, and more than 93\% of organizations are moderately or extremely concerned about cloud security. The challenge is how to improve cloud security and make data owners more confident in using public clouds. 

Researchers have been experimenting with novel crypto approaches, such as fully homomorphic encryption (FHE) \cite{brakerski11} and secure multi-party computation (SMC) \cite{huang11,mohassel17}, to address this problem. However, such pure software-based cryptographic solutions are still too expensive to be practical. Their overheads in storage, communication, and computation can be in many orders of magnitudes higher than the plaintext solutions.  Recent advances in hybrid protocols, e.g., for machine learning \cite{niko13sp,mohassel17,sharma19}, strive to reduce the overall costs of the frameworks by blending multiple crypto primitives to implement different algorithmic components. Although we have seen several close-to-practical solutions \cite{mohassel17,sharma19}, they are specifically designed for a particular problem or algorithm and difficult to generalize and optimize for new domains.  

During the past five years, the trusted execution environment (TEE) has emerged as a more efficient approach to addressing the performance and usability issues in secure outsourced computation. It provides hardware support to create an isolated environment within the cloud server that cannot even be compromised by adversaries who control the entire system software stack, including the operating system. Intel Software Guard Extension (SGX) \cite{sgx-explained} is probably the most well-known TEE hardware and SDK implementation. Since 2015, SGX has been available in most Intel CPUs. Using SGX, a user can run their sensitive computations in a TEE called \emph{enclave}, which uses a hardware-assisted mechanism to preserve the privacy and integrity of enclave memory. With SGX, users can pass encrypted data into the enclave, decrypt it, compute with plain text data, encrypt the result, and return it to the untrusted cloud components. 

SGX aims to protect the confidentiality and integrity of the in-enclave data, code, and computation.    
While it provides much better performance than expensive pure software-based crypto approaches, it leaks interactions and access patterns between the untrusted area and the enclave. Recent studies \cite{zheng17,dinh15,sasy18} show that some data access patterns may reveal application-level information to adversaries, even though the data blocks on the untrusted memory are encrypted. Thus, several efforts \cite{sasy18,ahmad18} adopt the concept of oblivious RAM (ORAM)  \cite{oded96} to address the access-pattern leakage problem. While ORAM is a generic solution covering all types of SGX applications by protecting each I/O access, it incurs high costs. With the state of the art ORAM schemes so far \cite{stefanov18,wang15}, each block access still comes with an additional $O(\log N)$ overhead, with $N$ numbers of blocks in the pool. Furthermore, for data-intensive applications, the data to be processed is normally larger than the limited physical enclave memory (i.e., the Enclave Page Cache) \cite{sgx-explained}. Thus, page swapping happens frequently, which can be precisely captured by adversaries via observing page fault interrupts \cite{shinde16} to infer the access pattern of sensitive data. 

\subsection{Scope of Our Research}
While ORAM provides a generic primitive for hiding data access patterns for all I/O operations, we find that it is unnecessary to hide sequential, uniform, and completely random I/O patterns, especially for data-intensive applications. For example, sequentially reading or writing a block file does not reveal additional information about the file content. However, the ORAM based I/O interface does not distinguish the different types of application patterns to treat them differently. Developers can certainly carefully examine each step in their applications to decide to use (or not to use) the ORAM operation. However, it is time-consuming and error-prone. Developers who are not familiar with the access-pattern attacks may accidentally mishandle or simply overlook some I/O accesses. We hypothesize that for many data-intensive algorithms, e.g., data mining algorithms, \emph{regulating the dataflow of the application and focusing on protecting specific access patterns} can be more efficient than using the ORAM primitive for all I/O accesses. We propose a framework: SGX-MR to automatically regulate the data flow of data-intensive applications and protect access patterns without depending on the expensive ORAM primitive or scrupulous examination and design by developers. The proposed approach protects not only the data access patterns from the untrusted memory but also record-level access patterns within the enclave threatened by page-fault attacks \cite{shinde16}. SGX-MR provides all of these benefits while keeping the low-level details transparent to the developers. 

Our idea is to take the MapReduce framework \cite{dean04} to regulate data-intensive SGX application dataflows, and then examine the stages of MapReduce processing to address their access pattern leakages. This approach has several advantages. (1) Since the dataflow in MapReduce processing is fixed for all MapReduce applications, it is more efficient for us to analyze each stage's access pattern, identify possible access-pattern leakages, and apply more efficient oblivious methods than ORAM. Using this systematic approach, it is possible to cover a large number of potential applications.  
(2) The MapReduce framework is easy to use and has been applied to numerous data-intensive problems, e.g., data mining \cite{miner12}, machine learning \cite{chu06}, text processing \cite{lin10}, and basic statistics. Researchers and practitioners have accumulated extensive experience \cite{chu06,miner12,lyubimov16} for solving most data analytics problems with the MapReduce framework during the past decade. Furthermore, Roy et al. \cite{roy10} also show that certain types of aggregate reducers with user-defined mapper functions will be sufficient to support a large class of data mining algorithms, including clustering, predictive modeling, and recommenders. SGX-MR strategically uses this finding to facilitate a wide range of data analytics while leveraging SGX’s features in solving the related confidentiality and integrity concerns. Therefore, SGX-MR\footnote{SGX-MR does not support the join operation yet, which will be examined for full performance and privacy guarantee in the future.} provides a versatile framework/tool for developers to implement wide range of data mining algorithms without having to worry about confidentiality and integrity concerns. Note that it is well known that applications requiring random accesses of data or low latency processing may not match the design purpose of MapReduce. Accordingly, SGX-MR will not be a good candidate for such applications.

We have carefully designed the whole SGX-MR framework\footnote{We will open-source the SGX-MR framework after we finalize the development.} to minimize the attack surface \cite{shih17} and achieve the best performance as possible. Specifically, our research has the following contributions.
\begin{itemize}
	\item We have implemented the lightweight SGX-MR framework that is flexible to adapt to the restricted enclave memory, data volume, and computation complexity. 
	\item We have carefully studied both the access patterns outside and inside the enclave during each stage of the SGX-MR framework and designed robust mitigation methods to prevent access pattern leakages.
	\item We have conducted extensive component-wise experiments to understand the cost and performance of the SGX-MR framework with the ORAM based SGX approach. The result shows that SGX-MR can be several times faster than ORAM based solutions.
\end{itemize}
In the remaining sections, we will first give the background knowledge for our approach (Section \ref{sec:preliminary}), then dive in the details of the proposed approach (Section \ref{sec:approach}), present the evaluation result (Section \ref{sec:eval}), and described the closely related work (Section \ref{sec:related}).
\section{Preliminary} \label{sec:preliminary}
We will present the related background knowledge before we dive into our approach, including the way SGX manages the enclave memory and related issues, the ORAM approach to addressing the known access-pattern attacks, and a brief introduction to MapReduce framework.   
\subsection{Enclave Memory Management}
Due to the hardware implementation cost, Intel SGX only reserves 128 MB of physical memory, known as Processor Reserved Memory (PRM). A certain portion of PRM has been used for maintaining SGX's internal operations. The remaining portion, about 90MB known as the Enclave Page Cache (EPC), can be used by the enclave programs. In the Linux systems, SGX can utilize the Linux virtual memory management for enclave programs, which can thus access a much larger virtual memory address space with contingency for potential page-faults. When a page-fault interrupt happens, the SGX driver will encrypt the selected EPC page and swap it out to the untrusted memory area. This memory management brings two concerns. First, since the SGX driver mostly uses the kernel's functions for page-fault handling, which an adversarial kernel can manipulate to issue targeted page-fault attack \cite{shinde16}. Shinde et al. \cite{shinde16} have shown such an attack can help adversaries learn the in-enclave access patterns over the EPC pages. Second, It is unknown how efficient the inherent SGX virtual memory management works in applications. Arnautov et al. \cite{arnautov16} show system-managed EPC page swapping incurs significant costs when handling data larger than the limit of enclave memory. We have also experimentally confirmed this observation. We also designed a block-buffer based application memory management to minimize the memory access pattern. However, due to the large processed data, the application-managed buffer also needs frequent swapping out blocks explicitly, showing no significant performance advantage over implicit system page swapping, as we show in experiments. Thus, for simplicity, a data-intensive application can just use the system virtual memory mechanism, e.g., by setting a sufficiently large enclave heap size. This also gives some flexibility for handling other issues like large objects.
\subsection{Access-Pattern Based Attacks and ORAM for SGX}
When encrypted messages are either accessed from memory or exchanged via the network, an adversary can observe the dataflow and possibly extract sensitive information. Cash et al. \cite{cash15} and Zhang et al. \cite{zhang16} have shown how an adversary can extract the content of encrypted documents by leveraging access pattern leakages. Ohrimenko et al. \cite{ohrimenko16} have also demonstrated how sensitive information, such as age-group, birthplace, marital status, etc., can be extracted from MapReduce programs by only observing the network flow and memory skew. 

While SGX can protect the content in the enclave memory from an adversarial OS, it does not provide a systematic mechanism to protect against access pattern leakages. When the program inside the enclave accesses the untrusted memory, the adversarial OS can always monitor the accessed memory addresses. Furthermore, adversaries can modify the virtual memory management mechanism to manipulate the page-fault interrupt generation to figure out in-enclave page access patterns \cite{shinde16}. 

Consider an example where a pharmaceutical company wants to find top sold items last month by analyzing the daily transactions in the cloud. As sales information is sensitive, they do not want to reveal any statistics from the transactions. Hence in a secure setting, data remains encrypted in untrusted memory and only decrypted in a trusted enclave during computation. A simple way is to use Hashmap to collect the counts of sold items: each item is mapped to a key, which is associated to a key-count pair located at a fixed memory address; each time an item is observed, the corresponding memory address is visited to update the corresponding count. Without access pattern protection, an adversarial OS can observe the memory page access pattern, which can be used to infer the number of a certain item accesses. This access pattern can be used to estimate the number of medicine types, the count of certain medicine, and the total number of sold items. 
\begin{figure}[t]
	\centering
	\centerline{\includegraphics[width=.8\linewidth]{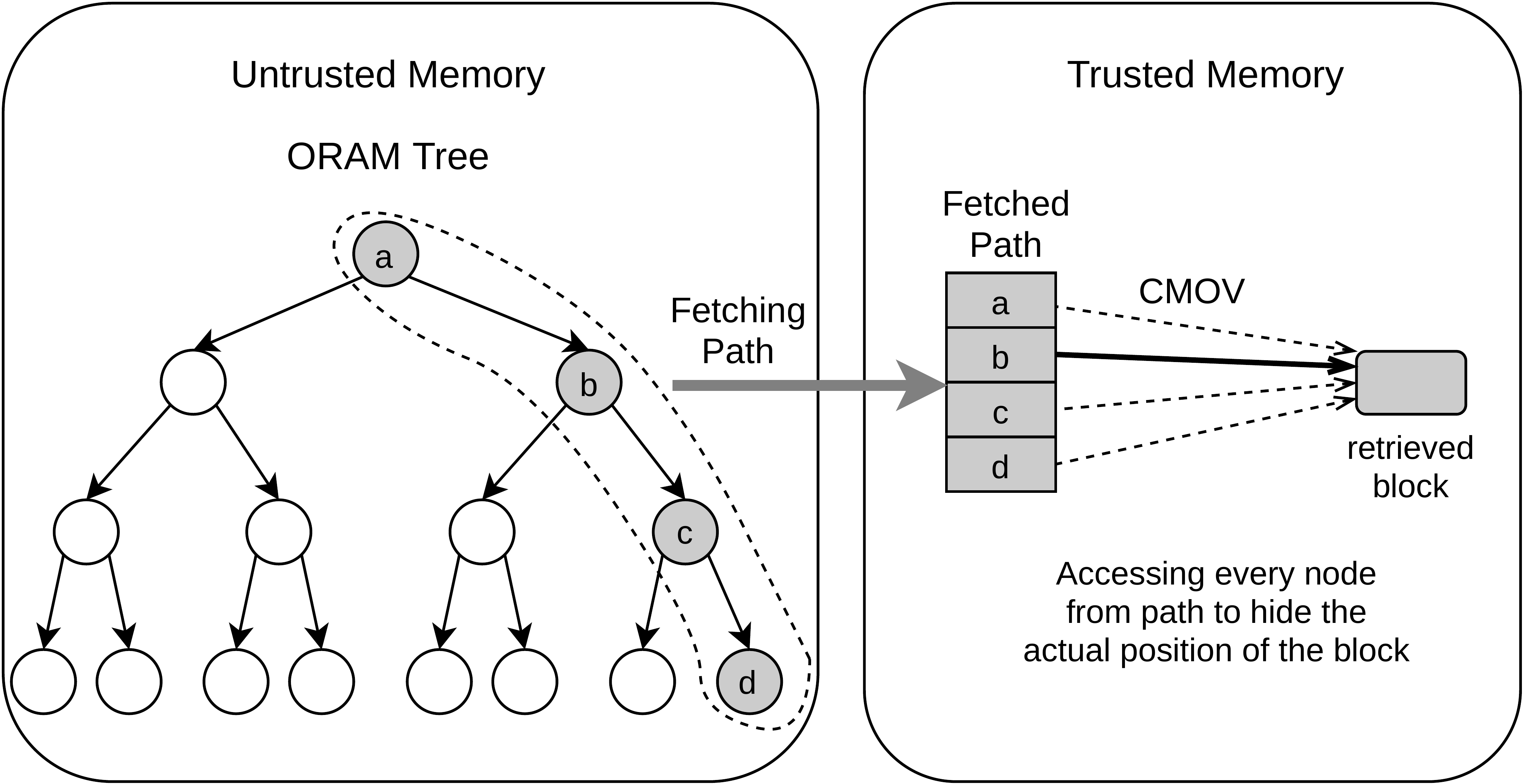}}
	\caption{Oblivious block retrieval from the ORAM Tree. The entire path containing the target block is loaded into the enclave to hide the access pattern. Furthermore, the CMOV instruction is used to hide the in-enclave access of the target block to address the page-fault attack.}
	\label{fig:oram-access}
\end{figure}

Researchers have applied the idea of Oblivious RAM (ORAM)\cite{oded96} to protect the block-level access patterns for SGX applications \cite{sasy18,ahmad18}.
ORAM primarily retrieves multiple memory blocks in a path and then randomly shuffles memory blocks periodically to disguise read/write access patterns on encrypted data blocks, where an adversary cannot learn additional information by observing memory access patterns. ORAM was originally developed for theoretical treatment for software protection\cite{oded96}, and then actively studied for protecting remote access patterns for cloud-based applications \cite{stefanov18,wang15} to address the privacy concerns with untrusted cloud service providers. By leveraging the ORAM construction, SGX applications can protect the access pattern of untrusted memory from the adversarial OS. The current SGX ORAM methods also partially address the in-enclave page-fault attack on ORAM-related operations. As shown by Figure \ref{fig:oram-access}, the popular SGX ORAM solutions, such as ZeroTrace \cite{sasy18} and Obliviate \cite{ahmad18}, utilize the most efficient Path ORAM \cite{stefanov18} or Circuit ORAM \cite{wang15}, which uses a tree structure and paths from the root to the leaves to hide the actual accessed block. The tree and data blocks are maintained in the untrusted area. Once the path containing the target block is identified, it is loaded into the enclave for the enclave program to extract the target block. However, adversaries may monitor page faults to figure out which block in the path is finally accessed. CMOV instruction\cite{olga16,rane15} is a CPU instruction that moves the source operand to destination if a conditional flag is set. Regardless the flag is set or not, the source operand will be read. Therefore, the occurrence of page fault cannot be used to infer whether the source is copied to the destination or not. A simplified example is shown as follows:     
\begin{verbatim}
//if (a < b) x = a else x = b
CMOVL x, a
CMOVGE x, b
\end{verbatim}
Consider $a$ and $b$ as two sources we want to protect (e.g., the blocks in the ORAM path, as shown in Figure \ref{fig:oram-access}). We want to hide the actual access pattern, say, copying $b$ to the buffer $x$. If the comparison result is true, the copy happens with the CMOVL (move if less) line, otherwise with the CMOVGE (move if greater or equal) line. For both lines, the sources $a$ and $b$ are read regardless of the condition is true or not.  
As the ORAM controller's vital data structures, such as the position map and stash, will be kept inside the enclave, the in-enclave access patterns about the position map and stash may also be under the page-fault attacks. Similar approaches have been used to hide these data-dependent page access patterns. However, as a lower level block I/O interface, SGX ORAM cannot address application-specific page-fault attacks.  
\subsection{MapReduce }
MapReduce\cite{dean04} is a popular programming model and also a processing framework, designed to handle large-scale data with a massively-parallel processing infrastructure. It has several major computation phases: map, optional combiner, shuffle and reduce. The input data is split into fixed-size blocks. Running in parallel, each Mapper takes the user-defined ``map'' function to process one assigned data block and converts it to key-value pairs. An optional combiner can be used to pre-aggregate the map outputs, if hierarchical aggregation works for the reduce function. Then, all output key-value pairs of mapper (or combiners if they are used) are sorted, grouped, and partitioned individually for Reducers to fetch. Each Reducer then fetches the corresponding share of Map (or Combiner) output in the shuffling phase. After collecting all shares, each Reducer sorts them. Finally, each Reducer applies the user-defined reduce function to process each group of key-value pairs to generate the final result. In both map and reduce phases, multiple processes (Mappers and Reducers) are run in parallel on distributed computing nodes to achieve high throughputs.
Noticing the regulated dataflow in MapReduce processing and the extensive experience in developing MapReduce applications accumulated during the past few years
\cite{chu06,miner12,lyubimov16}, we decide to apply this processing framework to the SGX-based data-intensive applications. By addressing the access pattern leakages in this framework, we can protect a large number of applications using our SGX-MR framework. 
\section{The SGX-MR Approach}\label{sec:approach}
Starting with an analysis of the typical data flow for data analytics algorithms in the SGX enclave context, we present the SGX-MR framework to integrate MapReduce processing into the SGX enclave environment. Then, we describe the target threat model and analyze possible access pattern leakages under the SGX-MR framework. We will design mitigation methods to address these access pattern leakages.
\subsection{Features of SGX-Based Data Analytics Algorithms and Threat Model}
Typical data analytics applications handle datasets much larger than the enclave memory and often they are sequentially scanned in multiple iterations.     
Using SGX for data analytics applications has several unique features. For simplicity, datasets are often organized in encrypted blocks and stored in a block-file structure. For security reasons, the SGX enclave program that runs in the protected EPC area cannot access the file system APIs directly. When processed, they are first loaded into the main (untrusted) memory, and then passed to the enclave via SGX APIs. Encrypted data blocks will be decrypted and processed inside the enclave. In this process, there are two security challenges. (1) Adversaries can observe the interactions between the untrusted memory and the enclave. Thus, preserving this access pattern is essential, which has been addressed by ORAM-based approaches, but appears too expensive. (2) Large data processing in enclave will inevitably cause data spill out, either handled implicitly by the system's page swapping mechanism, or explicitly by applications, e.g., via buffer management. Adversaries can also observe or even manipulate page faults to obtain in-enclave access patterns.

\textbf{Threat Model}.	
Based on the analysis of the features of SGX-based data analytics algorithms, we derive the following threat model. 
1)An adversary may compromise the operating system and any hypervisor of the host running the SGX applications. However, it cannot compromise the SGX hardware and enclaves. 2)The adversary may observe all of the data, program executions, and memory management of the system outside the SGX enclaves. It can thus analyze and detect interaction patterns between SGX enclave process and the untrusted memory. 3) A malicious adversary may tamper the application data and program running in the untrusted memory. 4) A powerful adversary may manipulate the data and program execution in the untrusted memory to force page faults within enclave to obtain in-enclave memory access patterns. 5) Denial-of-service and the side-channel attacks based on power analysis, or timing attacks \cite{hasindu18}, are out of the scope of this paper. 

We refer to the basic SGX mechanism \cite{sgx-explained} for integrity and confidentiality protection. We encrypt the code and data in the untrusted region of our framework with $sgx\_crypto$, a cryptographic library provided by Intel SGX SDK with robust security protection. We design the MapReduce framework and associated algorithms in such a manner that the page-fault attacks inside the enclave and the enclave's access patterns to the untrusted memory are oblivious from the adversary. 
\begin{figure}[t]
	\centering
	\centerline{\includegraphics[width=.8\linewidth]{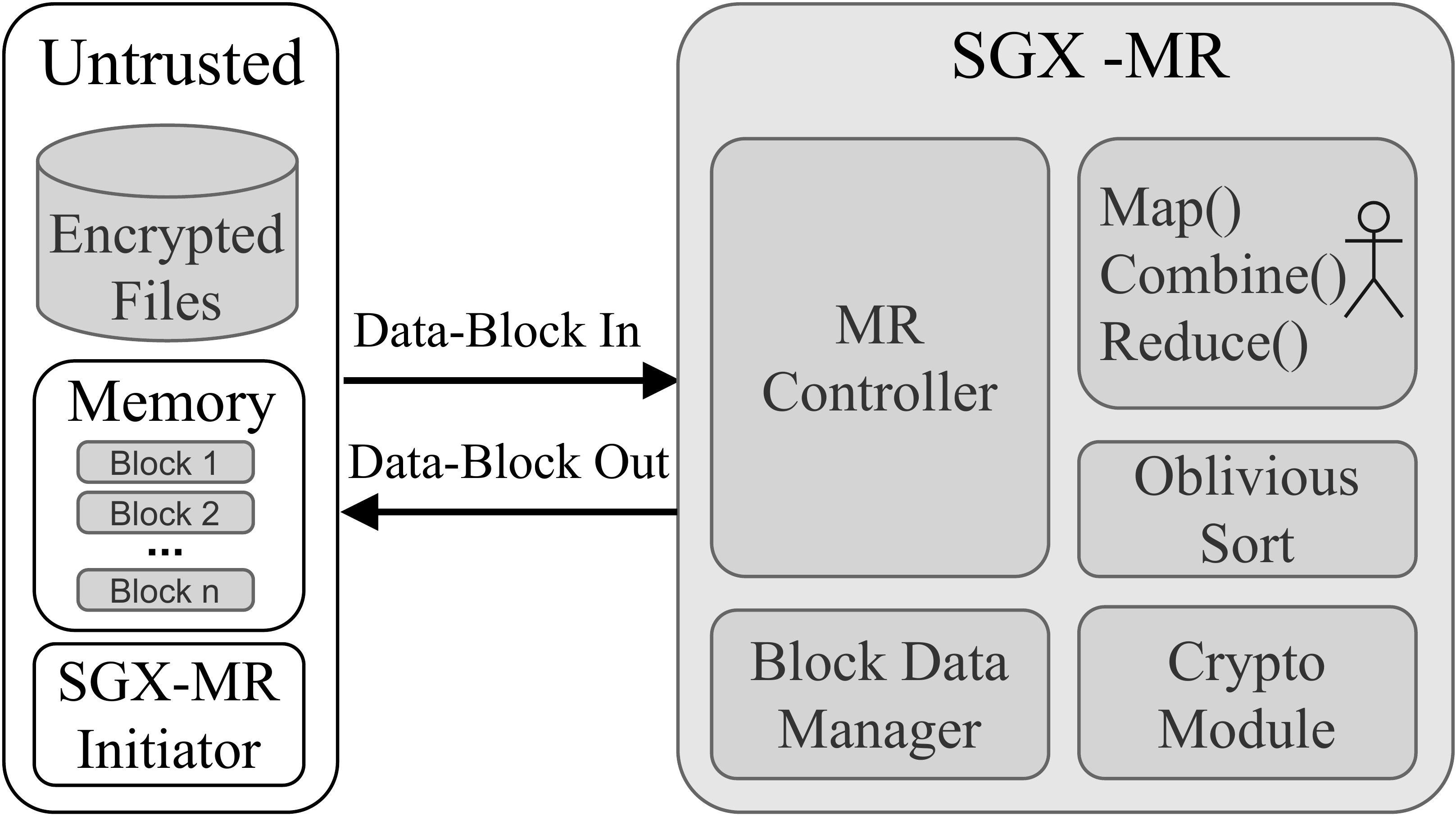}}
	\caption{High-level diagram of SGX-MR: shaded modules or memory either executed in the enclave or remain encrypted.}
	\label{fig:framework}
\end{figure}
\subsection{Design of SGX-MR }
According to the SGX working mechanism and features of data-intensive applications, we partition the entire framework into two parts, i.e., the trusted (enclave) and untrusted parts. Figure \ref{fig:framework} shows the components. The untrusted part contains the user-uploaded encrypted data and the small block I/O library. The remaining components of the SGX-MR framework reside in the enclave.
\begin{itemize}
	\item \textbf{Untrusted Part}. Since the data and the I/O library are in the untrusted area, our design needs to address both their confidentiality and integrity. We design a block file structure for encrypted data on disk and in the untrusted memory area. For simplicity, we assume that data records have a fixed size. Both block size and record size are tunable by the user based on the specific application. Each block also contains a message authentication code (MAC) to ensure data integrity. The whole block is encrypted systematically using the AES Counter (CTR) mode functions in the SGX SDK. To minimize the attack surface, we design the library running in the untrusted part to handle only block I/O, and a verification function inside the block data manager in the enclave to capture any adversarial modification on the loaded data blocks. 
	\item \textbf{Enclave Part}. The SGX-MR controller handles MapReduce jobs and controls the dataflow of the application. Users only provide the map(), reduce(), and combine() functions to implement the application-specific processing. For simplicity, we will focus on the aggregation-style combine and reduce functions, such as COUNT, MAX, MIN, SUM, TOP-K, etc., which have been shown sufficient to handle many data analytics tasks \cite{roy10}. Remarkably, with our careful design, the binary of the whole SGX-MR framework (without the application-specific map, combine, and reduce functions) takes only about 1.1MB physical memory. With the manually managed memory, we can also work with EPC memory as small as 3--4 blocks. The block size may depend on the specific application (we use block sizes varying from 2 KB to 2 MB for both WordCount and kMeans application in our experiment). 
\end{itemize}
\subsubsection{Dataflow Regularization in SGX-MR}\label{sec:sgx-mr-regulated}
\begin{figure}[t]
	\centerline{\includegraphics[width=.8\linewidth]{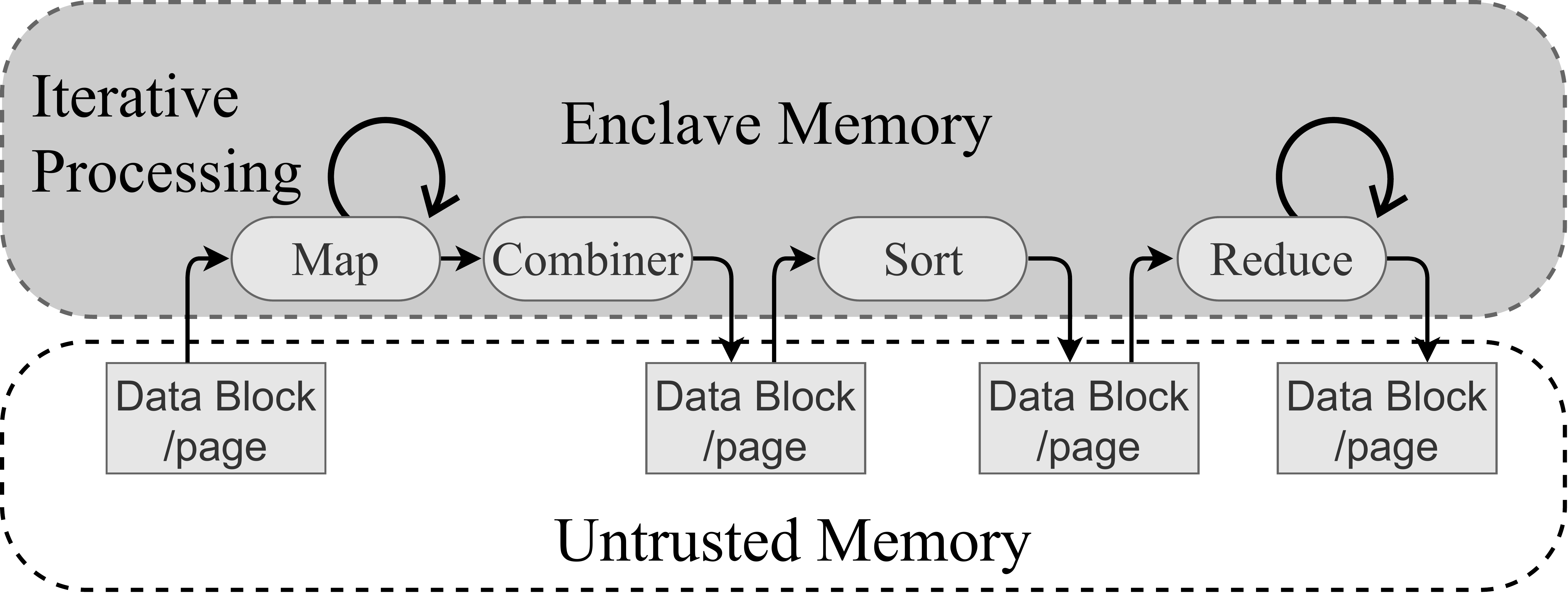}}
	\caption{Regulated dataflows between enclave and main memory}
	\label{fig:regulated-dataflow}
\end{figure}

With a basic understanding of the components in SGX-MR, we describe how the application dataflow is regulated, which helps simplify access-pattern analysis and protection. While the original MapReduce is designed for parallel processing, Figure \ref{fig:regulated-dataflow} sketches how the dataflow regulated by the MapReduce processing pipeline and processed sequentially in SGX-MR, and the interactions between the enclave and the untrusted memory. First of all, input files are processed by the data owner, encoded with the block format via a \emph{file encoding} utility tool, and uploaded to the target machine running the SGX-MR application. Second, within a MapReduce job, all file access requests from the enclave (i.e., mapper reading and reducer writing) have to go through the SGX-MR block I/O module running in the untrusted memory area. Third, the intermediate outputs, e.g., of the Map, Combiner, and Sorting phases, can also be spilled out either by application buffer manager or system's virtual memory manager, in encrypted form. 

Specifically, after the job starts, the Map module will read encrypted data blocks sequentially from the untrusted memory area, decrypt them, and apply the user-provided \texttt{map} function to process the records iteratively, which generates the output in key-value pairs. Note that the formats of both the input records and the generated key-value pairs are defined by users for the specific application (for readers who are not familiar with MapReduce programming, please refer to Section \ref{sec:preliminary} and the original paper \cite{dean04}). The controller accumulates the generated key-value records until they fill up a block, and then sort them by the key. With strictly managed memory, the filled data block will be written to the untrusted area temporarily. If we depend on the virtual memory management, the filled data blocks will stay in the enclave memory and swapped out by the system when needed. We have evaluated both options in experiments and found that application-managed memory has performance advantages. Since we have restricted the reducer functions to a set of hierarchically aggregatable functions (i.e., aggregates can be done locally and then globally), we also design their combiner functions for local aggregation. For example, for the COUNT function, the combiner will generate the local counts for a key, say $k_i$: $( (k_i, c_1), (k_i, c_2), \dots, (k_i, c_m)) $ if there are $m$ mappers. The reduce phase will get the final counts of $\sum_{j=1}^m c_j$. The inclusion of combiner has two benefits: (1) it can significantly reduce the cost of the most expensive phase: sorting, and (2) it can indirectly address the group size leakage problem in the reduce phase that will be discussed later. 

The Combiner outputs will go through the sorting phase, where the sorting algorithm will sort all key-value pairs into groups uniquely identified by the key. The Reduce phase then iteratively processed the groups of key-value pairs identified by the key. 

Like the Map phase, key-value pairs stored in blocks will be handled sequentially in the Reduce phase. Specifically, the user-selected \emph{reduce} function from the library takes a group of records with the same key and generates aggregated results. For all the aggregation functions we mentioned earlier, a sequential scan over the group will be sufficient to generate the aggregates. The aggregate of each group is also in the form of key-value pairs, which are accumulated in data blocks, encrypted, and written back to the untrusted area. 
The above described dataflow keeps the same for all applications that can be cast into the MapReduce processing framework. Now, by specifically addressing the potential access-pattern leakages in the MapReduce data flow, we can effectively protect a broad category of data-intensive SGX applications from access pattern related attacks.
\subsubsection{Block Design}
We fix the record size and records per block to avoid variable length records leaking additional information. Active adversaries can observe the record-level access pattern via page-fault attacks, which allows them to identify record length. Fixed-length records can effectively address this problem, with a cost of padding for short records. One may concern about the efficiency of handling extra-large record that exceeds the limit of the physical enclave memory (e.g., $>100$ MB) . We admit the padding will take a significant extra cost, but the limit of enclave memory may not be a concern. By adjusting the enclave heap size to sufficiently accommodate blocks, the system's virtual memory management can handle well --- in this case, we prefer not using the application-managed buffer. Experiments have shown the performance difference between different memory management strategies is ignorable. 
\subsubsection{Integrity Guarantee}
While SGX assures the integrity of enclave memory, both code and data that reside in untrusted memory remain vulnerable and can be modified. SGX-MR minimizes untrusted execution that was only used for storing and retrieving block data from untrusted memory. The integrity of the untrusted execution will be verified inside the enclave. 

We consider three possible attacks to integrity: (1) modify a block, (2) shuffle a block with another block in the same file, and (3) insert a block from a different file (or a phase's output that is encrypted with the same key). To address all the attacks, we include the following attributes in the block: (i) Block ID, so that block shuffling can be identified, (ii) File Id, so that no block from different files can be inserted, and (iii) the block-level Message Authentication Code (MAC). At the end of each block, a MAC is attached to guarantee the integrity of records, before the whole block is encrypted. We also use the randomized encryption using AES-CTR encoding to make sure the same block will be encrypted to a totally non-distinguishable one so that adversaries cannot trace the generated results in the MapReduce workflow. A simple verification program runs inside the enclave that verifies the IDs and MAC, after reading and decrypting a block.  
\section{Access-Pattern Leakages and Mitigation Methods}
Based on the SGX-MR dataflow analysis, we can identify several critical data access patterns pertaining to different phases: Map's input, intermediate processing, combining, and output; Shuffling/Sorting's input, sorting, and output; and Reduce's input, aggregation, and output. Among these access patterns, Map's input and Reduce's output involve only sequential block reads/writes. Thus, individually they do not leak additional information except for input and output file sizes (i.e., the number of blocks). However, Reducer's output, if observed synchronously with its input, may reveal some information as we show in the following. In this section we examine the access pattern leakages in different stages of SGX-MR and discuss the mitigation methods. First, we described leakage in sorting, followed by leakages by page faults, and finally the leakages in the reducing phase of the framework.
\subsection{Leakage in Sorting }
\begin{figure}[t]
	\centerline{\includegraphics[width=.8\linewidth]{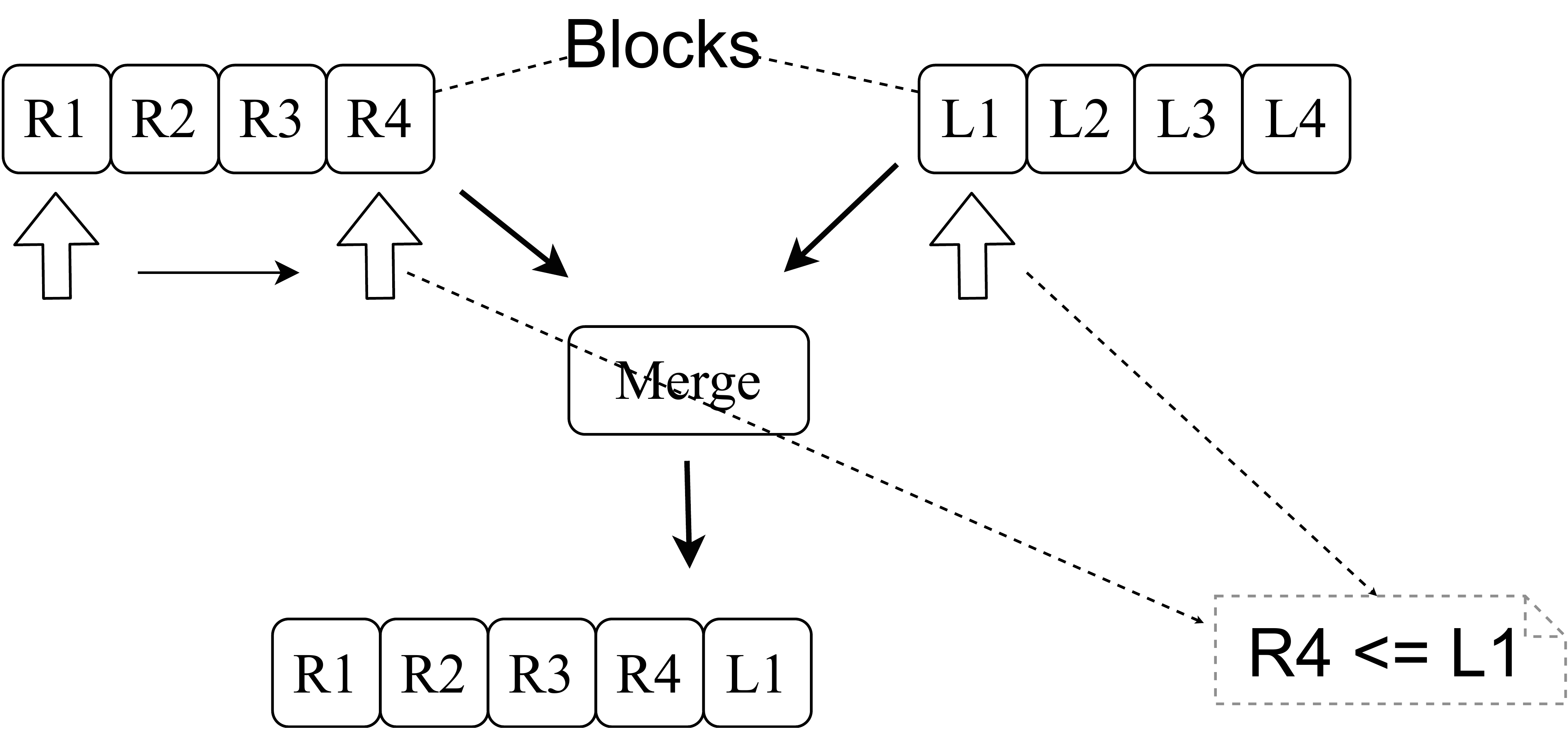}}
	\caption{Access-pattern leakage in MergeSort by observing the interactions between the enclave program and the untrusted memory. If R4 $\leq$ L1, then the Merge Phase reads R1 to R4 one by one to write sorted output and then reads L1 to L4. By observing this reading pattern (movement of the pointers), the adversary will learn that the entire right sub-list is smaller than the left sub-list. 
	}
	\label{fig:sorting-leakage}
\end{figure}

We start with the most expensive part of the whole data flow - the Sorting phase. Traditionally, the MapReduce framework adopts the MergeSort algorithm for its simplicity. As the shares of Map(or combiner)-output have been sorted individually, MergeSort only needs to merge the sorted shares. However, as we show next, it leaks vital information about the records under sorting. 

If we look at the block-wise merge process carefully, we can identify some unusual merging behaviors, which reveals some sensitive attributes of the processed records. Starting from the records in each block sorted by the Map phase, the main body of MergeSort is to merge and sort two individually sorted lists of records recursively until all records are sorted. Assume the two merged lists contain the records $\{L_i\}$ and $\{R_j\}$, respectively, as shown in Figure \ref{fig:sorting-leakage}. If one of the lists has all the values larger (or less) than the other, the corresponding access pattern will be continuously reading the blocks in one list first, followed by the whole list of the other. This pattern may leak sensitive information for real applications. For example, in the WordCount program, adversaries might be able to guess the frequencies of words and derive the word distribution. 

\begin{figure}[h]
	\centerline{\includegraphics[width=.8\linewidth]{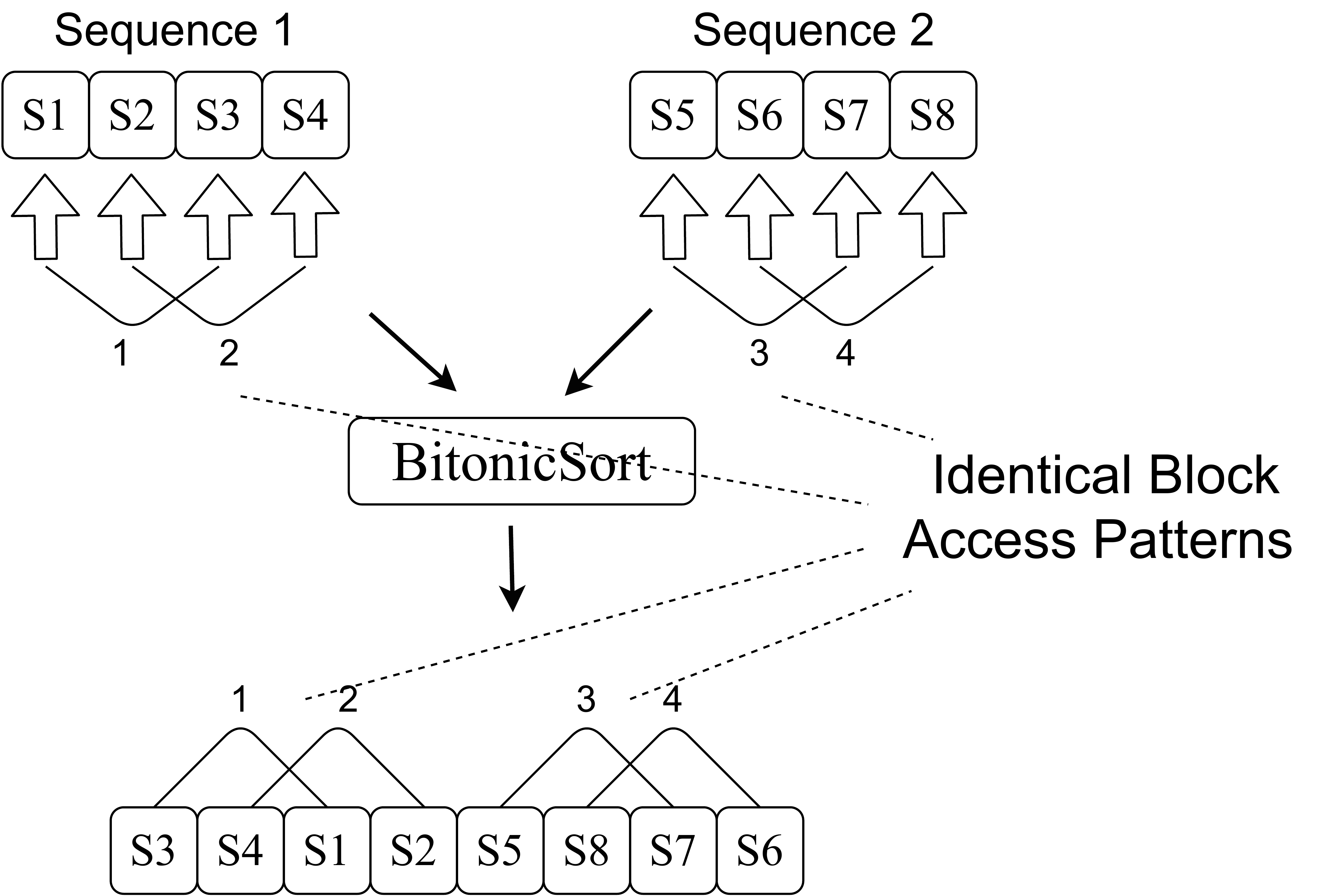}}
	\caption{Uniform block access does not leak any pattern in Bitonic Sort which is oblivious to the data in the blocks. The figure shows the state of the sorting algorithm in the second iteration of sorting}
	\label{fig:bitonic-no_leakage}
\end{figure}

\textbf{Mitigation Methods.} This access pattern problem involves two parts: block-level pattern in untrusted memory and the page-level pattern in enclave memory. In the following, we address the block-level patterns only with \emph{oblivious sorting} algorithms \cite{batcher68}. The in-enclave page-level patterns will be discussed later. By definition, regardless of the actual values in the sorted list, an oblivious sorting algorithm will take the fixed identical access pattern that is only determined by the size of the list. In the SGX-MR framework, we implement the well-known \emph{BitonicSort} \cite{batcher68} algorithm (See Fig \ref{fig:bitonic-no_leakage}. Note that BitonicSort takes $O(N(\log N)^2)$ block accesses for $N$ blocks, compared to MergeSort's cost $O(N\log N)$. Our evaluation shows although it is more expensive than MergeSort, the overall cost is still significantly less than MergeSort with ORAM-based oblivious I/O interface. 
\subsubsection{In-Enclave Page-Access-Pattern Leakage}

As previously discussed, enclave execution is vulnerable to the page-fault attack \cite{shinde16}. Since the in-enclave processing is around records, we look at the related record processing. First, the page-fault attack does not provide additional information for sequential record accesses in executing the map function and the reduce aggregation function. However, we do find some non-sequential in-enclave access patterns that need to be protected from page-fault attacks. They include the in-enclave sorting part of the Sorting phase and the map-output sorting (when a block is filled up) before combining. The repetitive page accesses for the same data block reveal the ordering of records in a pair of blocks, a similar scenario to Figure \ref{fig:sorting-leakage}, but happening at the page level.

\textbf{Mitigation Methods.} At first glance, we can just use the BitonicSort algorithm to hide the record-level access pattern (as a result of page-level access leakage) inside the enclave. However, this is insufficient, as the core operation of this in-enclave BitonicSort, \emph{compare-and-swap}, can still be captured by observing the sequence of page faults and used to reveal the relative ordering between pair of records. 
The following code snippet shows how possibly the access pattern is associated with the record order. 
\begin{verbatim}
if (a >= b){
// swap a and b, and 
// the page access can be observed.
}else{
// no page access.
}
\end{verbatim}

To avoid this access-pattern leakage, we adopt the oblivious swap operation applied to different scenarios \cite{constable18}. The basic idea is to use the CMOV instructions to hide the page access patterns. The oblivious swap operation indeed occurs high cost. We have observed about $2-2.5\times$ cost increase for finishing the Sorting phase. 
\subsection{Leakage in Reducing} 
\label{sec:leakage-in-reducer}
\begin{figure}[t]
	\centerline{\includegraphics[width=.8\linewidth]{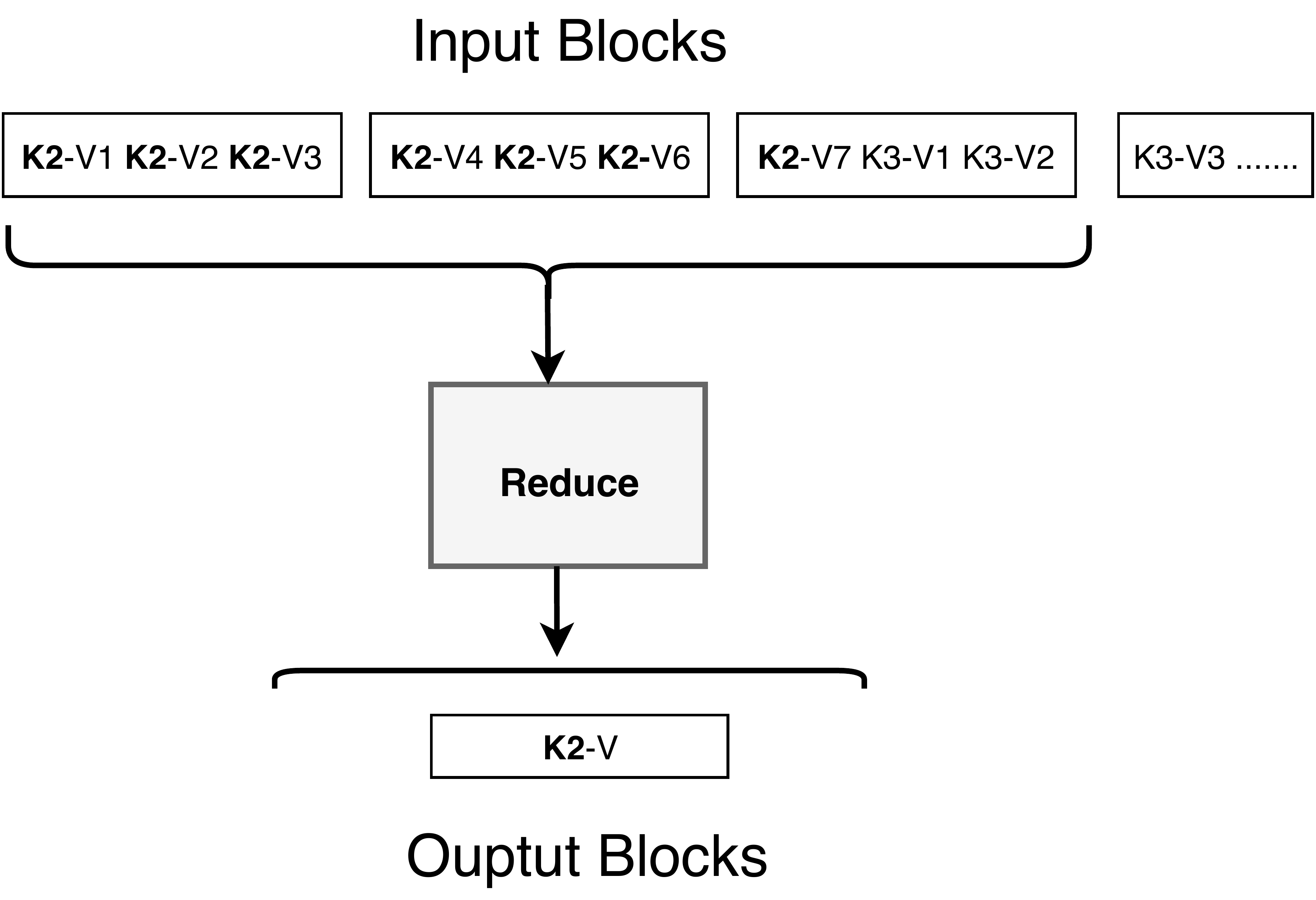}}
	\caption{Group-wise aggregation in Reduce phase may leak group sizes. K2 spread over three blocks. Thus, the access pattern will be reading three blocks sequentially and then possibly writing out one block. 
	}
	\label{fig:group-size-leakage}
\end{figure}

The Sorting phase will order key-value pairs by the key. As a result, all pairs will be sorted into groups. The controller will sequentially read the sorted blocks and transfer the records containing the same key (i.e., a group of records) to the reduce function. For aggregation-based reduce functions, each group will be reduced to one key-value pair. By observing the input/output ratio, adversaries may estimate group sizes, as shown in Figure \ref{fig:group-size-leakage} (if Combiners are not presented). In extreme cases, multiple blocks can be reduced into one record, which may lead to severe information leakage. For example, it leaks the word frequencies for the WordCount program, which helps estimate the word distribution that, in turn, can be used to guess words. 

\textbf{Mitigation Methods.} We consider a simple method to address the group-size leakage problem. As described in the previous section, the size difference between the Reducer's input and output leaks the group sizes. However, as we have made the Combiner component mandatory in SGX-MR thanks to the restricted types of reduce functions, it helps  obfuscate the aggregated result in the reduce phase. For example, without Combiners, each record in the reduce phase corresponds to one occurrence of a word (i.e., a key) in the WordCount program, and the group size reveals the frequency of the word. With Combiners, each record entering the Reduce phase may correspond to the pre-aggregate of multiple records. Thus, counting the records in the Reducer input does not give a good estimation of actual group sizes. Note that we do not protect the combining process from page-fault attacks (e.g., the CMOV-based method) for achieving better performance, which will possibly expose partial group sizes. However, adversaries cannot trace back the Reducer input to the Combiner output to utilize the partial group size information because the oblivious sorting phase between them breaks the link entirely. A concern is whether the \emph{frequency} of such combiner-output records still preserves the \emph{ranking} of actual group sizes. The intuition is that the key of a large group may be presented in more map outputs than a key of a small group.  We analyzed the observed group size after adding combiners to check whether the size ranking is till preserved. Figure \ref{fig:word-freq} for WordCount and Figure \ref{fig:cluster-size} for kMeans both show the ranking of group sizes is not preserved at all. 
To completely address the possible ranking leakage, we can further take one of the two candidate approaches: (1) injecting dummy key-value pairs by Combiners, e.g., adding (key, 0) for non-presented keys in the Combiner output for the COUNT function, or (2) disguising the reduce patterns by padding the output block with dummy records, e.g., every time an input block is scanned, output a padded block, regardless of whether or not the actual aggregation happens. In the following, we describe the second method in more detail.
\begin{verbatim}
load a block;
  for each record in the block
    if (the end of the group)
      write the aggregate record
    else
      write a dummy record
write the output block
\end{verbatim}

The above process shows the dummy record writing process. Note that all the records will be encrypted and thus non-distinguishable. The concern is that the page-fault attack can catch the execution of the if-else branch, which leaks the information of writing aggregate or dummy record. To address this problem, we also apply the CMOV protection here.
Another problem is the increased size of output due to the added dummy records, which can be a significant cost for applications like kMeans that have only a few aggregated records generated. We apply an additional bitonic sorting over the padded blocks and extract top-k elements to get rid of the dummy records, so that the group size ranking is still oblivious to adversaries. In experiments, we will evaluate the additional costs associated with this method.

\subsection{Maintaining the Integrity of the Specifications}

The IEEEtran class file is used to format your paper and style the text. All margins, 
column widths, line spaces, and text fonts are prescribed; please do not 
alter them. You may note peculiarities. For example, the head margin
measures proportionately more than is customary. This measurement 
and others are deliberate, using specifications that anticipate your paper 
as one part of the entire proceedings, and not as an independent document. 
Please do not revise any of the current designations.

	\section{Experimental Evaluation}\label{sec:eval}
\subsection{Experiment Setup}

SGX-MR is implemented with C++ and the Intel SGX SDK for the Linux environment. Our core framework consists of only about 2000 lines of code. The entire SGX-MR framework runs inside the enclave, except a small component in the untrusted area that handles block-level read/write requests from the enclave. The compiled framework without the application code takes about 1.1MB. To match the design specification, we have implemented a customized bitonic sort that works with block-level accesses in the untrusted memory. To protect record level access patterns in the enclave, we also apply the bitonic merge operation to obliviously merge in-enclave blocks. Furthermore, we hide the conditional swaps and sensitive branching pattern by leveraging CMOV instructions. As described in section 4.2, we have implemented mandatory combiners to obfuscate the actual group sizes for the selected aggregators. 
The experiments were conducted on a Linux machine with an Intel(R) Core(TM) i7-8700K CPU of 3.70GHz processor and 16 GB of DRAM. We use ZeroTrace's open-source implementation for our experiments. We use 128-bit AES-CTR encryption to encrypt the data blocks in the untrusted memory.

\textbf{Sample Applications.} We use two sample applications in our evaluation: WordCount and KMeans. WordCount takes a document collection and output the frequency for each word. It is an essential tool for language modeling and has been included in various tutorials as an example of data-intensive processing. KMeans \cite{jain99} is a fast, simple clustering algorithm. It takes the initial cluster centroids and iteratively conducts the two steps: (1) cluster membership assignment for each record and (2) the centroid re-computation, until the clustering converges, where the centroids (or all records' cluster membership) do not change anymore. In all KMeans related experiments, we only execute one iteration of this learning process.

\textbf{Baseline Implementation.}
To compare SGX-MR with ORAM-based approaches, we have implemented ORAM-based sample applications without depending on the MapReduce framework. In most data analytics programs, aggregating by groups is one of the most expensive core operations. It can be implemented with either hashing or sorting (as in the MapReduce workflow). With ORAM based access-pattern protection, one may wonder whether hashing can be more efficient for enclave-based programs. However, processing hash table in enclave creates page-level pattern \cite{kim19,oblix}. For example, to update a hash value, we must access the position of the particular hash key and update its value list. An adversary can observe this access pattern and possibly figure out the group size that ORAM cannot help. A few recent studies have noticed the problems with enclave based hashtable. Taehoon et al. \cite{kim19} propose a key-value data-structure for extensive data, but it still leaks the access pattern of the encrypted keys on a hash table. Mishra et al. \cite{oblix} propose an ORAM based key-value data structure to fully address the access pattern leakage. However, it takes $O(N)$ complexity to find a key in a worst-case scenario, and it will take $O(N^2)$ computational cost to find the aggregated results. On the other hand, MergeSort with ORAM and CMOV takes only $O(Nlog^2 N)$ to sort the records, and a consequent linear scan is sufficient to compute the aggregated results. Thus, in our evaluation, we will take MergeSort with CMOV protection for ORAM based baseline implementations to efficiently address both block and page level attacks. 
\subsection{Performance of Core Operations}
\begin{figure}[t]
	\centering
	\begin{tikzpicture}
	\begin{axis}[
	width=(\linewidth)*0.9,
	line width=1pt,
	scaled ticks=false,
	xlabel={Number of Blocks},
	ylabel style={align=center},
	ylabel={Execution Time(ms)},
	xmin=3000, xmax=21000,
	ymin=1, ymax=550000,
	xtick={3000,9000, 15000, 21000},
	ytick={100000,10000, 1000, 100, 10},
	legend pos=north west,
	ymajorgrids=true,
	grid style=dashed,
	ymode=log,
	]
	
	\addplot[
	cyan,
	mark=*,
	/tikz/dashed,
	]
	coordinates {
		(3000,2508.06)
		(6000,5488.64)
		(9000,8918.06)
		(12000,11858.59)
		(15000,14796.98)
		(18000,19218.09)
		(21000,22412.64)
	};
	\addlegendentry{ORAM Block Access}
	
	\addplot[
	color=red,
	mark= *,
	]
	coordinates {
		(3000,9.26)
		(6000,18.45)
		(9000,27.84)
		(12000,36.8)
		(15000,46.54)
		(18000,55.82)
		(21000,64.44)
	};
	\addlegendentry{Sequential Access in SGX-MR}
	
	\end{axis}
	\end{tikzpicture}
	
	\caption{For sequential access over data ORAM incurs significant cost overhead compared to SGX-MR. 
	}\label{fig:blockaccess}
\end{figure}
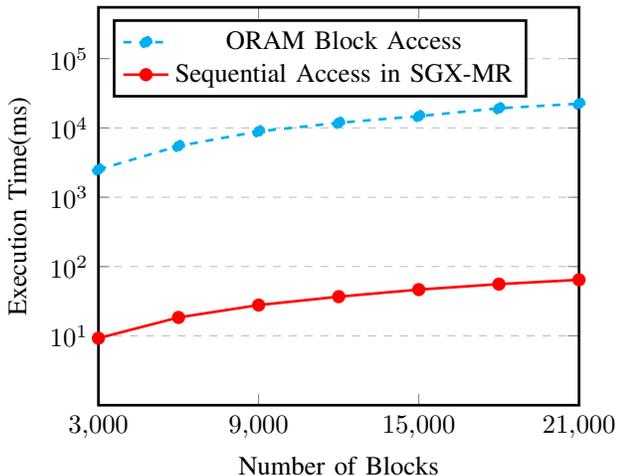%
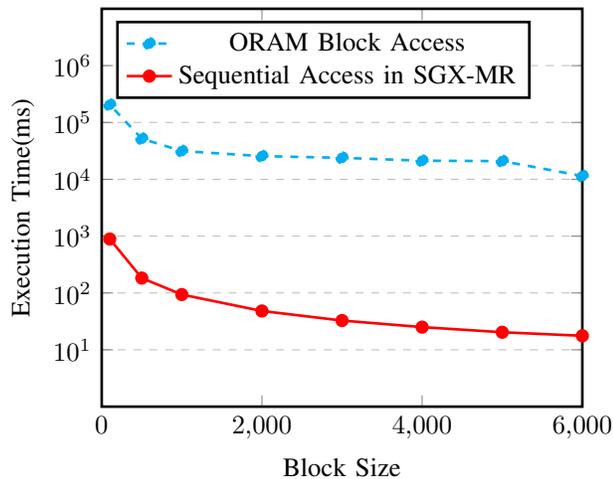
\begin{figure}[t]
	\centering
	\begin{tikzpicture}
	\begin{axis}[
	width=(\linewidth)*0.9,
	line width=1pt,
	scaled ticks=false,
	xlabel={Block Size},
	ylabel style={align=center},
	ylabel={Execution Time(ms)},
	xmin=0, xmax=6000,
	ymin=1, ymax=10000000,
	xtick={0,2000,4000,6000},
	ytick={1000000, 100000,10000,1000,100,10},
	legend pos=north west,
	ymajorgrids=true,
	grid style=dashed,
	ymode=log,
	]
	
	\addplot[
	cyan,
	mark=*,
	/tikz/dashed,
	]
	coordinates {
		(6000,11434.92)
		(5000,20877.22)
		(4000,21216.15)
		(3000,23849.72)
		(2000,25522.93)
		(1000,31391.57)
		(500,51802.71)
		(100,205278.64)
	};
	\addlegendentry{ORAM Block Access}
	
	\addplot[
	color=red,
	mark= *,
	]
	coordinates {
		(6000,17.65)
		(5000,20.36)
		(4000,25.07)
		(3000,32.7)
		(2000,48.18)
		(1000,93.73)
		(500,182.97)
		(100,890.14)
	};
	\addlegendentry{Sequential Access in SGX-MR}
	\end{axis}
	\end{tikzpicture}
	\caption{Effect of different block size on ORAM and SGX-MR sequential access, with a fixed file 30 MB.} \label{fig:blockaccess-blocksize}
\end{figure}%

\textbf{Block Access.} We started our evaluation with the basic operations to understand the sources of performance gain for SGX-MR compared to ORAM. As sequential block accesses do not provide valuable access-pattern information to adversaries, ORAM protection is unnecessary. Figure \ref{fig:blockaccess} shows that ORAM protected block accesses can be 1000x slower than those without protection. As SGX-MR's input, output, and some intermediate steps, are sequential block accesses, we can utilize this feature to save a significant amount of cost. 
Since block size is a critical application-specific factor, we measured the impact of block size on performance. Figure \ref{fig:blockaccess-blocksize} shows that a larger block size can significantly reduce the overhead of transferring the same amount of data (30 MB) into enclave. While the block size is increased from 1 KB to 6 MB, the execution time is reduced by about 50 to 100 times for both ORAM and sequential access. Since the ORAM cost is affected by the number of blocks, a larger block size reduces the number of blocks and thus also benefits ORAM.

\textbf{Sorting.} Another key operation in our framework is key-value record sorting. We have shown that it is necessary to use oblivious sorting algorithms to protect the important relationship between keys. We use MergeSort in the ORAM setting and BitonicSort for SGX-MR in the comparison. In both solutions, we also used CMOV-based protection for page-fault attacks. Figure \ref{fig:sorting} shows that BitonicSort under SGX-MR is much faster than MergeSort with ORAM for the block I/O. While the costs of both algorithms are asymptotically the same, i.e., $O(n\log^2{n})$, MergeSort+ORAM appears to have a much high constant factor ---  it is about five times slower when sorting 20,000 blocks. 
\begin{figure}[t]
	\begin{tikzpicture}
	\begin{axis}[
	width=(\linewidth)*0.9,
	line width=1pt,
	scaled ticks=false,
	xlabel={Number of Blocks},
	ylabel={Execution Time (ms)},
	xmin=0, xmax=20000,
	ymin=0, ymax=700000,
	xtick={0,5000,10000,15000,20000},
	ytick={100000, 200000, 300000, 400000, 500000, 600000, 700000},
	legend pos=north west,
	ymajorgrids=true,
	grid style=dashed,
	]
	\addplot[
	cyan,
	mark=*,
	/tikz/dashed
	]
	coordinates {
		(3000,61630.29)(4000,81994.25)(5000,121446.31)(6000,145477.56)(8000,194424.28)(10000,284250.85)(12000,340200.36)(15000,427489.26)(18000,589199.45)(20000,656385.49)
	};
	\addlegendentry{Merge Sort + ORAM}
	\addplot[
	color=red,
	mark= *,
	]
	coordinates {
		(3000,8428.384)(4000,8583.388)(5000,19483.325)(6000,19660.032)(8000,20065.591)(10000,45120.083)(12000,45567.785)(15000,46049.07)(18000,102917.335)(20000,103247.565)
	};
	\addlegendentry{Bitonic Sort in SGX-MR}
	\end{axis}
	\end{tikzpicture}
	\caption{MergeSort with the ORAM block I/O results in significantly higher costs than a dedicated oblivious sorting (e.g., BitonicSort) in SGX-MR. Note: block size = 1 KB. 
	}
	\label{fig:sorting}
\end{figure}
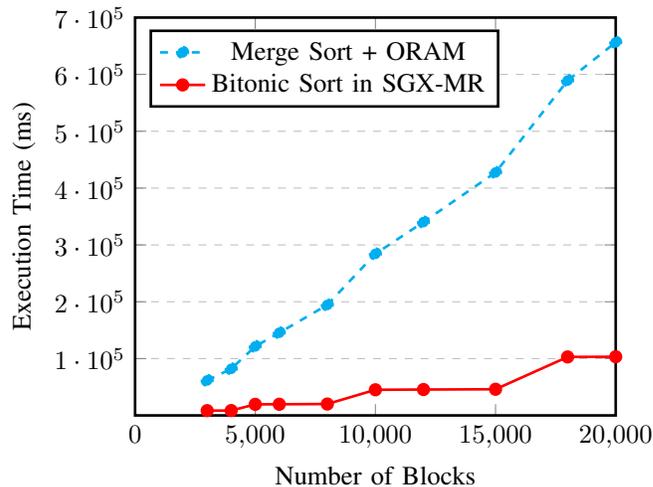
\begin{figure}[t]
	\centering
	\begin{tikzpicture}
	\begin{axis}[
	width=0.9*(\linewidth),
	title style={align=center},
	line width=1pt,
	scaled ticks=false,
	legend style={nodes={scale=1, transform shape}},
	xlabel={Allocated Memory (megabyte)},
	ylabel={Execution Time ($\mu$s)},
	xmin=0, xmax=120,
	ymin=0, ymax=8,
	xtick={20, 40, 60, 80, 100},
	ytick={1, 2, 3, 4, 5},
	legend pos=north west,
	ymajorgrids=true,
	grid style=dashed,
	]
	\addplot[
	color=red,
	mark= x,
	]
	coordinates {
		(10,0.18)
		(20,0.228)
		(30,0.251)
		(40,0.264)
		(50,0.27)
		(60,0.276)
		(70,0.281)
		(80,0.727)
		(90,2.222)
		(100,3.405)
		(110,4.316)
	};
	\addlegendentry{Random Access}
	\addplot[
	cyan,
	mark=square*,
	]
	coordinates {
		(60,0.002)
		(70,0.002)
		(80,0.006)
		(90,0.006)
		(100,0.006)
		(110,0.006)
	};
	\addlegendentry{Sequential Access}
	\draw[dashed, gray] (90,0) -- (90,5) node[above=3pt]{EPC Limit}; 
	\end{axis}
	\end{tikzpicture}
	\caption{When allocated buffer size go beyond EPC limit, system-managed memory incurs significant cost overhead for random data access. Conversely, sequential access has negligible impact on it.
	}\label{fig:epcmemory}
\end{figure}
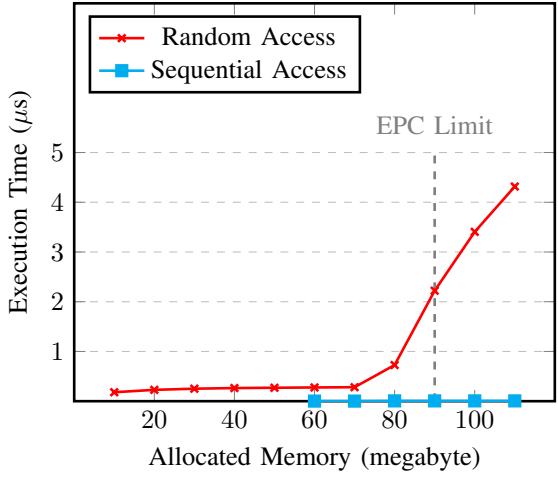%

\textbf{Impact of Enclave Memory Management.}
The physical memory allocated for SGX enclaves, called Enclave Page Cache (EPC), is limited to a small size --- enclave applications can use only about 90 MB. However, the SGX Linux library can utilize the Linux virtual memory management so that applications can access the large virtual memory space. Thus, application developers have two options: either entirely depending on the enclave virtual memory management, or manually managing a limited data buffer to minimize the system intervention. Sergei et al. \cite{arnautov16} noticed that when the data size is small manually managed buffer can cost less than the enclave virtual memory management as it can reduce the number of unnecessary page swaps. We revisited this issue and show that for sufficiently large datasets (e.g., > 90 MB) these two strategies make not much difference. 
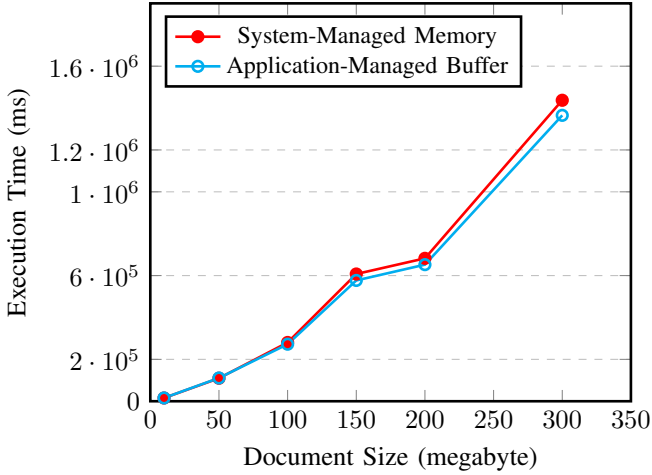
\begin{figure}[t]
	\centering
	\begin{tikzpicture}
	\begin{axis}[
	width=0.9*(\linewidth),
	legend style={nodes={scale=1.3, transform shape}}, 
	title style={align=center},
	line width=1pt,
	scaled ticks=false,
	legend style={nodes={scale=0.7, transform shape}},
	xlabel={Document Size (megabyte)},
	ylabel={Execution Time (ms)},
	xmin=0, xmax=350,
	ymin=0, ymax=1900000,
	xtick={0,50,100, 150, 200,250, 300, 350},
	ytick={0,200000,600000,1000000,1200000, 1600000 },
	legend pos=north west,
	ymajorgrids=true,
	grid style=dashed,
	]
	\addplot[
	color=red,
	mark= *,
	]
	coordinates {
		(10,15390.569)
		(50,109752.949)
		(100,280584.698)
		(150,607527.096)
		(200,682364.908)
		(300,1437238.332)
	};
	\addlegendentry{System-Managed Memory}
	\addplot[
	cyan,
	mark=o
	]
	coordinates {
		(10,15621.016)
		(50,111019.593)
		(100,272378.951)
		(150,576838.757)
		(200,652241.403)
		(300,1365238.759)
	};
	\addlegendentry{Application-Managed Buffer}
	\end{axis}
	\end{tikzpicture}
	\caption{BitonicSort in SGX-MR costs for different memory management strategies. System-managed memory incurs small cost overhead compared to application-managed Buffer.
	}\label{fig:epcmemory-sort}
\end{figure}%
Figure \ref{fig:epcmemory} confirms that page-fault processing takes a significant amount of time. We conduct random block accesses over an array passed to the enclave, which is copied to the enclave heap. When the array size exceeds the limit of EPC size, the EPC page fault handling mechanism is frequently called for random accesses, which incurs higher costs. 
However, when working with large data, swapping EPC memory by either buffer management or the system paging does not make much cost different. Figure \ref{fig:epcmemory-sort} compares BitonicSort with a four-block buffer management strategy, verses with all blocks loaded into a big array and managed by the system. The cost difference is negligible. However, sometimes the application managed buffer might be desired, e.g., to minimize the memory footprint or better control memory swapping.  
\subsection{Costs of Access-Pattern Protection in SGX-MR}

We have discussed several mechanisms to protect the specific access patterns in SGX-MR. In this section, we design a set of experiments to fully understand how these costs contribute to the entire processing costs. 

\textbf{Oblivious Sorting.}
Figure \ref{fig:sorting-cost} shows the cost comparison with access pattern protection (i.e., BitonicSort) vs. without (i.e., MergeSort) for the WordCount algorithm implemented with SGX-MR. SGX-MR with BitonicSort is approximately seven times slower than SGX-MR with MergeSort. Thus, this cost is significant. 
\begin{figure}[h]
	\begin{tikzpicture}
	\begin{axis}[
	width=(\linewidth)*0.9,
	title style={align=center},
	line width=1pt,
	scaled ticks=false,
	legend style={nodes={scale=0.7, transform shape}},
	xlabel={Document Collection Size (megabytes)},
	ylabel={Execution Time (ms)},
	xmin=0, xmax=200,
	ymin=0, ymax=13000000,
	xtick={0, 50, 100, 150, 200},
	ytick={3000000,6000000,9000000,12000000},
	legend pos=north west,
	ymajorgrids=true,
	grid style=dashed,
	]
	\addplot[
	color=cyan,
	mark=square*,
	]
	coordinates {
		(10,22422.445)
		(20,60216.864)
		(50,385259.673)
		(100,958394.107)
		(150,979754.371)
		(200,2333619.397)
	};
	\addlegendentry{SGX-MR with BitonicSort}
	\addplot[
	color=red,
	mark=o,
	]
	coordinates {
		(10,  128563.32)
		(20,  333121.851)
		(50,2018872.408)
		(100,4976940.598)
		(150,5086943.945)
		(200,12014268.35)
	};
	\addlegendentry{SGX-MR with BitonicSort + o-swap}    
	\addplot[
	color=orange,
	mark= *,
	]
	coordinates {
		(10,9688.933)
		(20,21661.958)
		(50,66425.588)
		(100,146019.228)
		(150,219093.947)
		(200,316336.428)
	};
	\addlegendentry{SGX-MR with MergeSort}
	\end{axis}
	\end{tikzpicture}
	\caption{SGX-MR with BitonicSort vs. with MergeSort. O-swap uses CMOV to protect from page-fault attacks. Note: WordCount with block size = 2 MB}
	\label{fig:sorting-cost}
\end{figure}
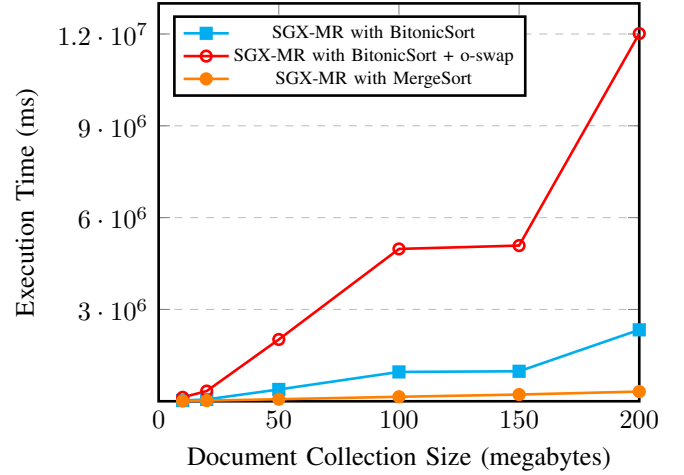
\begin{figure}[htb]
	\begin{tikzpicture}
	\begin{axis}[
	width = \linewidth * 0.9,
	title style={align=center},
	legend style={nodes={scale=.8, transform shape}}, 
	line width=1pt,
	scaled ticks=false,
	xlabel={Block Size},
	ylabel={Execution Time (ms)},
	xmin=500, xmax=6000,
	ymin=20000, ymax=1000000,
	xtick={500, 2000, 4000,6000},
	ytick={100000, 300000,500000, 700000, 900000},
	legend pos=north east,
	ymajorgrids=true,
	grid style=dashed,
	]
	\addplot[
	color=red,
	mark=*,
	/tikz/dashed
	]
	coordinates {
		(6000,283888.47)
		(5000,284099.22)
		(4000,342145.38)
		(3000,359735.78)
		(2000,438763.03)
		(1000,577723.71)
		(500,883289.38)
	};
	\addlegendentry{MergeSort + ORAM}
	\addplot[
	color=orange,
	mark= *,
	]
	coordinates {
		(6000,38264.529)
		(5000,33894.861)
		(4000,58559.333)
		(3000,48223.717)
		(2000,75334.643)
		(1000,103143.517)
		(500,147594.457)
	};
	\addlegendentry{SGX-MR: BitonicSort}
	\addplot[
	color=cyan,
	mark=square,
	]
	coordinates {
		(6000,  142947.334)
		(5000,  140024.145)
		(4000,  162380.821)
		(3000,  156254.313)
		(2000,  179277.406)
		(1000,  203901.734)
		(500, 244430.402)
	};
	\addlegendentry{SGX-MR: BitonicSort with o-swap}
	\end{axis}
	\end{tikzpicture}
	\caption{Effect of block size on the cost of ORAM+MergeSort vs. BitonicSort with/without o-swap. 
	}
	\label{bitonicsort-block}
\end{figure}
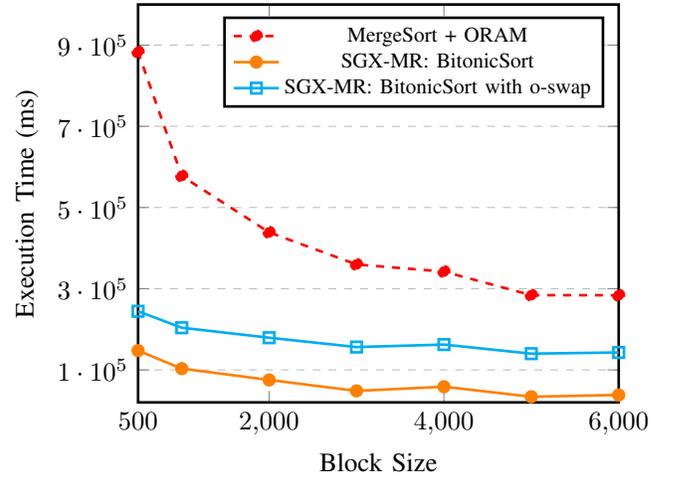

\textbf{In-enclave Page-Fault Attack protection.}
Next, we evaluate the protection of the page-fault attack on oblivious sorting. Previously, we have shown that protecting access patterns from untrusted memory is not enough. Even in enclave memory, page-level access patterns reveal the order of the records in an oblivious sorting algorithm.  As described in section \ref{fig:sorting-leakage}, we applied the oblivious swap technique to protect the order of the records within each block. 

Fig. \ref{fig:sorting-cost} and \ref{bitonicsort-block}  show that,  compared to ORAM+MergeSort, the additional cost brought by BitonicSort+o-swap is relatively small. 
\begin{figure}[htp]
	\centering
	\begin{tikzpicture}
	\begin{axis}[
	width=(\linewidth)*0.9,
	line width=1pt,
	scaled ticks=false,
	xlabel={Top 100 words ordered by frequency},
	ylabel={Word Frequency},
	xmin=0, xmax=100,
	ymin=1, ymax=1000000,
	xtick={0, 20, 40, 60, 80, 100},
	ytick={1, 10, 100, 1000, 10000, 100000},
	legend pos=north west,
	ymajorgrids=true,
	grid style=dashed,
	ymode=log,
	log basis y={10}
	]
	
	\addplot[
	color=red,
	]
	coordinates {
		(0, 11841)
		(1, 9809)
		(2, 8291)
		(3, 7681)
		(4, 7198)
		(5, 6956)
		(6, 5069)
		(7, 4023)
		(8, 3493)
		(9, 3361)
		(10, 3334)
		(11, 3067)
		(12, 2821)
		(13, 2743)
		(14, 2688)
		(15, 2669)
		(16, 2629)
		(17, 2594)
		(18, 2577)
		(19, 2576)
		(20, 2560)
		(21, 2557)
		(22, 2425)
		(23, 2343)
		(24, 2322)
		(25, 2085)
		(26, 1994)
		(27, 1992)
		(28, 1938)
		(29, 1922)
		(30, 1849)
		(31, 1834)
		(32, 1762)
		(33, 1744)
		(34, 1622)
		(35, 1589)
		(36, 1524)
		(37, 1487)
		(38, 1447)
		(39, 1415)
		(40, 1264)
		(41, 1258)
		(42, 1233)
		(43, 1226)
		(44, 1213)
		(45, 1208)
		(46, 1041)
		(47, 1036)
		(48, 988)
		(49, 921)
		(50, 921)
		(51, 897)
		(52, 883)
		(53, 875)
		(54, 869)
		(55, 864)
		(56, 829)
		(57, 814)
		(58, 790)
		(59, 782)
		(60, 778)
		(61, 771)
		(62, 769)
		(63, 764)
		(64, 747)
		(65, 744)
		(66, 737)
		(67, 719)
		(68, 714)
		(69, 708)
		(70, 693)
		(71, 689)
		(72, 679)
		(73, 658)
		(74, 650)
		(75, 642)
		(76, 628)
		(77, 625)
		(78, 614)
		(79, 597)
		(80, 590)
		(81, 588)
		(82, 584)
		(83, 553)
		(84, 552)
		(85, 538)
		(86, 522)
		(87, 500)
		(88, 499)
		(89, 498)
		(90, 496)
		(91, 488)
		(92, 481)
		(93, 481)
		(94, 476)
		(95, 473)
		(96, 473)
		(97, 453)
		(98, 449)
		(99, 446)
	};
	\addlegendentry{Actual Frequency}
	
	\addplot[
	color=cyan,
	]
	coordinates {
		(0, 7)
		(1, 1)
		(2, 1168)
		(3, 1)
		(4, 8)
		(5, 1)
		(6, 4)
		(7, 1)
		(8, 1)
		(9, 1)
		(10, 6)
		(11, 1)
		(12, 1)
		(13, 85)
		(14, 1)
		(15, 2)
		(16, 1)
		(17, 3)
		(18, 2)
		(19, 1)
		(20, 1)
		(21, 1)
		(22, 1)
		(23, 109)
		(24, 3)
		(25, 1)
		(26, 892)
		(27, 2)
		(28, 1)
		(29, 1)
		(30, 1)
		(31, 3)
		(32, 100)
		(33, 1)
		(34, 6)
		(35, 7)
		(36, 1)
		(37, 17)
		(38, 4)
		(39, 1)
		(40, 1)
		(41, 1)
		(42, 19)
		(43, 114)
		(44, 1)
		(45, 5)
		(46, 1)
		(47, 1)
		(48, 1)
		(49, 2)
		(50, 1)
		(51, 1)
		(52, 1)
		(53, 14)
		(54, 2)
		(55, 4)
		(56, 1)
		(57, 26)
		(58, 480)
		(59, 10)
		(60, 1)
		(61, 4)
		(62, 6)
		(63, 17)
		(64, 1)
		(65, 1)
		(66, 2)
		(67, 1)
		(68, 1)
		(69, 516)
		(70, 35)
		(71, 10)
		(72, 1)
		(73, 481)
		(74, 2)
		(75, 446)
		(76, 2)
		(77, 2)
		(78, 3)
		(79, 1)
		(80, 112)
		(81, 1)
		(82, 29)
		(83, 1)
		(84, 23)
		(85, 1)
		(86, 11)
		(87, 1)
		(88, 1)
		(89, 1)
		(90, 1)
		(91, 1)
		(92, 1)
		(93, 11)
		(94, 1)
		(95, 371)
		(96, 2)
		(97, 4)
		(98, 373)
		(99, 1)
	};
	\addlegendentry{Observed Frequency}
	
	\end{axis}
	\end{tikzpicture}
	\caption{Observed group sizes before (exactly mapped to word frequencies) and after adding combiners for WordCount. 
	}
	\label{fig:word-freq}
\end{figure}
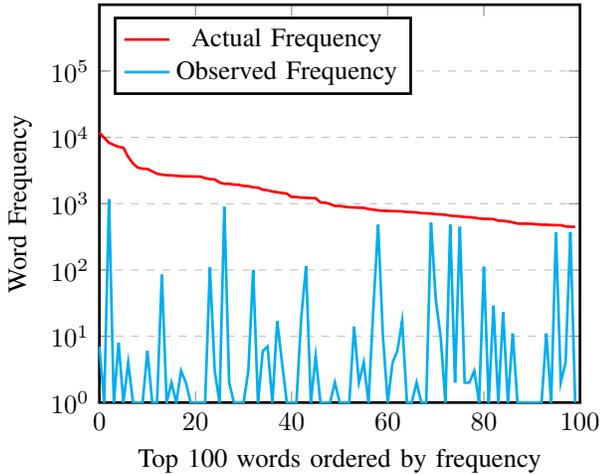
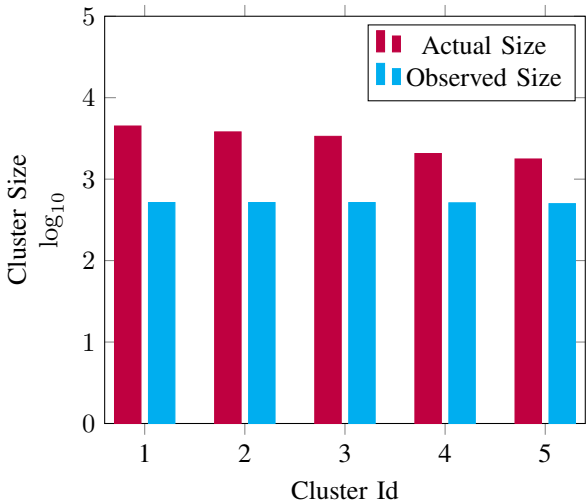
\begin{figure}[htb]
	\begin{tikzpicture}
	\begin{axis}[
	ybar=.1cm,
	bar width=10pt,
	ymin=0,
	ymax=5,
	height=7cm,
	width=\linewidth * .9,
	xticklabels={1, 2, 3, 4, 5},
	ylabel style={align=center},
	ylabel={Cluster Size \\ $\log_{10}$},
	xlabel={Cluster Id},
	xtick=data,
	]
	\addlegendentry{Actual Size}            
	\addlegendentry{Observed Size}            
	\addplot [draw=purple, fill=purple]  coordinates {(1, 3.6521496054) (2,3.5798978696) (3, 3.52543355343) (4,3.31428866095) (5,3.24748226068)};
	\addplot [draw=cyan, fill=cyan] coordinates {(1, 2.71264970163) (2, 2.71264970163) (3, 2.71264970163) (4,2.70926996098) (5, 2.69897000434) };
	\end{axis}
	\end{tikzpicture}
	\caption{ Observed group sizes before (exactly mapped to the	cluster size) and after adding combiners for kMeans.}
	\label{fig:cluster-size}
\end{figure}

\textbf{Protecting Group Sizes in Reducing.} We have mentioned that by using Combiners mandatorily in our framework we can effectively protect from the group-size-estimation attack in reducing. Figure \ref{fig:word-freq} and \ref{fig:cluster-size} how Combiners' outputs disguise the actual group sizes. It’s well-known that combiners improve performance as they can significantly reduce the number of records going to the Reduce phase. We achieved about $\times 2$ and $\times 7$ speedup for WordCount and KMeans respectively. As our experiments showed, combiners are especially beneficial for applications with a small number of keys such as KMeans. To further protect the possibly preserved group size ranking, we have also evaluated the dummy-record padding method. Table \ref{tab:padding} shows the storage cost of dummy record padding method and its post-processing cost to remove dummy records. The padding will significantly increase the storage cost. However, with the post-processing step, which approximately doubles the processing time of the Reduce phase, the size goes back to the original one, which will not affect the performance of future processing.      

\captionsetup{
	justification = centering
}
\begin{table*}[h!]
	\begin{center}
		\centering
		\caption{Additional Costs by Padding}
		\label{tab:padding}
		\begin{tabular}{l|l|l|l|l|l}
			\textbf{Application} & \textbf{Input} & \textbf{Actual Output} & \textbf{Padded Output} & \textbf{w/o post-processing} & \textbf{with post-processing}\\
			& block & block & block & ms & ms \\
			\hline
			KMeans & 10,000 & 1 & 1,662 & 33146 & 51606\\ 
			\hline
			
			WordCount & 10,000 & 3,997 & 53,439 & 1022055 &2029600\\
			
		\end{tabular}
	\end{center}
\end{table*}
\subsection{Application-Based Evaluation}
Finally, we compare the SGX-MR and ORAM implementations for the sample applications. The SGX-MR WordCount compiled code has around 1.6MB and the KMeans code has around 1.5MB. Samples of the Yelp review dataset\footnote{https://www.kaggle.com/yelp-dataset/yelp-dataset } are used for evaluating the WordCount algorithm. The KMeans algorithm is evaluated with a simulated two-dimensional dataset that fills 30 records per 2 KB block and up to 10,000 blocks. We implement the ORAM-based solutions with the original algorithms (not in the MapReduce style) and use the ZeroTrace block I/O interface for accessing data. As we discussed earlier, the ORAM-based baselines use an o-swap protected MergeSort for aggregating the records. 
Figure \ref{fig:wordcount} and \ref{fig:kmeans} show the overall costs according to the increased data. Overall, the SGX-MR-based implementations for both applications are significantly faster than ORAM-based implementations (Table \ref{tab:application-bar}). 
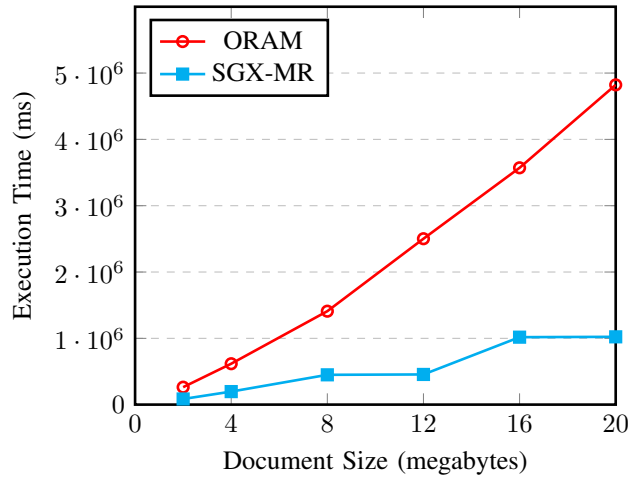
\begin{figure}[htb]
	\begin{tikzpicture}
	\begin{axis}[
	width=(\linewidth)*0.9,
	line width=1pt,
	scaled ticks=false,
	xlabel={Document Size (megabytes)},
	ylabel={Execution Time (ms)},
	xmin=0, xmax=20,
	ymin=1000, ymax=6000000,
	xtick={0,4,8,12,16,20},
	ytick={0, 1000000, 2000000, 3000000, 4000000, 5000000},
	legend pos=north west,
	ymajorgrids=true,
	grid style=dashed,
	]
	\addplot[
	color=red,
	mark=o
	]
	coordinates {
		(2,263820.1)
		(4,617533.19)
		(8,1409558.36)
		(12,2502221.62)
		(16,3571523.16)
		(20,4821916.211)
	};
	\addlegendentry{ORAM}
	\addplot[
	color=cyan,
	mark= square*,
	]
	coordinates {
		(2,85757.773000)
		(4,197243.046000)
		(8,450129.107000)
		(12,455706.218)
		(16,1017746.348)
		(20,1022974.149)
	};
	\addlegendentry{SGX-MR}
	\end{axis}
	\end{tikzpicture}
	\caption{Application Level Comparison of WordCount Problem
	}
	\label{fig:wordcount}
\end{figure}
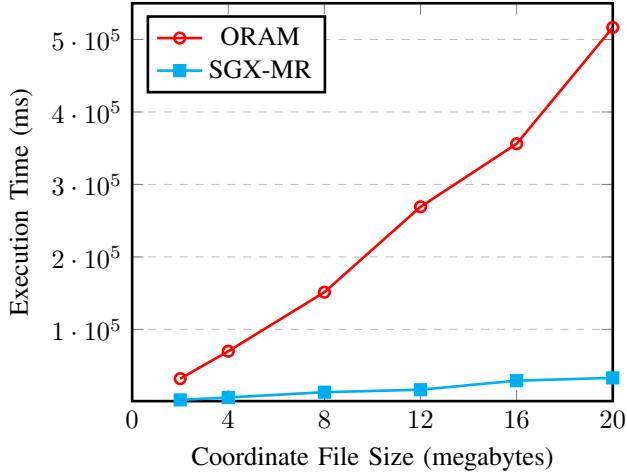
\begin{figure}[htb]
	\begin{tikzpicture}
	\begin{axis}[
	width=(\linewidth)*0.9,
	title={Application Level Comparison of KMeans Problem},
	line width=1pt,
	scaled ticks=false,
	xlabel={Coordinate File Size (megabytes)},
	ylabel={Execution Time (ms)},
	xmin=0, xmax=20,
	ymin=1000, ymax=550000,
	xtick={0, 4, 8, 12, 16, 20},
	ytick={100000, 200000, 300000, 400000, 500000},
	legend pos=north west,
	ymajorgrids=true,
	grid style=dashed,
	]
	\addplot[
	color=red,
	mark=o,
	]
	coordinates {
		(2,32026.31)
		(4,69848.02)
		(8,151298.35)
		(12,269121.78)
		(16,356145.03)
		(20,516634.11)
	};
	\addlegendentry{ORAM}
	\addplot[
	cyan,
	mark= square*,
	]
	coordinates {
		(2,2708.263)
		(4,6019.486)
		(8,13335.774)
		(12,16746.528)
		(16,29209.189)
		(20,33078.161000)
	};
	\addlegendentry{SGX-MR}
	\end{axis}
	\end{tikzpicture}
	\caption{Application Level Comparison of KMeans Problem
	}
	\label{fig:kmeans}
\end{figure}
\begin{table}[h!]
	\begin{center}
		\caption{Application Level Comparisons of SGX-MR and ORAM for different applications. Block size = 2 KB, data size = 20 MB.}
		\label{tab:application-bar}
		\begin{tabular}{l|l|l|l}
			\textbf{Application} & \textbf{SGX-MR} & \textbf{ORAM} \\
			& (ms) & (ms) \\
			\hline
			KMeans & 33078 & 516634\\ 
			\hline
			
			WordCount & 1022974 & 4821916\\
			
		\end{tabular}
	\end{center}
\end{table}
\section{Related Work} \label{sec:related}
There are several pieces of work related to our study. The most popular technique for access-pattern protection would be Oblivious RAM \cite{oded96}. During the past years, the research on ORAM \cite{stefanov18,wang15} has been booming due to the rise of cloud computing and the privacy concerns over cloud data store. Tree-based Path ORAM \cite{stefanov18} and Circuit ORAM \cite{wang15} are among the most efficient schemes. ZeroTrace \cite{sasy18} uses both Path and Circuit ORAM to address the access-pattern problem for SGX applications. While using the most efficient implementation of ORAM, we show that ZeroTrace based block I/O still takes a significant overhead over the regular block I/O. Obliviate \cite{ahmad18} also uses the Path ORAM scheme to design the I/O interface for an SGX-based secure file system. Similarly, the additional cost is significant. Both schemes try to design a protection mechanism at the block I/O level, providing a high level of transparency to application developers. However, we argue that an application-framework-level protection mechanism can be more efficient.

Researchers also try to extend big data processing platforms to take advantage of SGX. The basic idea is to keep the codebase of current software, such as Hadoop \cite{dean04} and Spark \cite{zaharia10}, unchanged as possible, while moving the data-processing parts to the SGX enclave. VC3 \cite{schuster15} applied this strategy for modifying the Hadoop system. It moves the execution of ``map'' and ``reduce'' functions to the SGX enclave, while the upper-level functions such as job scheduling and data management still stay outside the enclave. The most part of the Hadoop Java library is not changed at all. As a result, it achieves the goal of processing encrypted sensitive data in enclaves, but leaves other issues, such as access pattern protection and computation integrity (for the components running in the untrusted memory area), not addressed. 

M2R \cite{dinh15} addresses the problem of access-pattern leakage in the shuffling phase of VC3 and proposes to use the oblivious schemes for shuffling. However, other security problems are still not addressed.
These top-down approaches have the fundamental problem --- unless the whole framework is re-implemented and moved to the enclave, adversaries can easily attack the components running in the untrusted area. Our work is entirely different from these studies. We utilize the MapReduce to regulate the application dataflow so that we can apply access-pattern protection mechanisms to the framework level. 
Opaque \cite{zheng17} tries to revise Spark for SGX. They also focus on the data access patterns between computing nodes and illustrate how adversaries can utilize these access patterns to infer sensitive information in the encrypted data. To protect the distributed data access patterns, they provide four types of primitive oblivious operators to support the rich Spark processing functionalities: oblivious sort, filter, join, and aggregation. Noticing the problem of computation integrity, they try to move the job controller part, formerly in the master node, to the trusted client-side and design an integrity verification method to detect whether worker nodes process data honestly. However, to reuse the most parts of Spark codebase, it has the most of the system running in the untrusted area, especially for worker nodes. Thus, the local-level integrity guarantee and access-pattern protection might be insufficient. Mostly, it shares the same fundamental problem with VC3 and M2R, as we mentioned. 

\section{Conclusion}
	ORAM has been a popular solution for protecting access-pattern leakages in SGX-based applications. Yet, it is still very expensive. Furthermore, since ORAM provides a protection mechanism for lower-level  block I/O, it cannot address the application-level access-pattern leakage. We also notice that application-specific data access patterns can take advantage of alternative oblivious access techniques to improve performance. Thus, we propose the SGX-MR framework to regulate the dataflow with the MapReduce processing framework, which works for a large class of data-intensive applications. The regulated dataflow allows us to analyze the access pattern leakages at each stage and develop solutions for all applications implemented with the SGX-MR framework. We have conducted extensive experiments to understand the features of the SGX-MR. The result shows that SGX-MR-based implementations of data-intensive applications work much more efficiently than ORAM-based implementations. We will continue to optimize the phases of the SGX-MR framework to achieve even better performance.

\bibliographystyle{abbrv}
\bibliography{./paper}

\end{document}